\documentclass[doublecol]{epl2} 

\usepackage{amsmath}
\usepackage{amssymb}
\usepackage{graphicx}
\usepackage{epsfig}

\title{Forced flow of granular media: Breakdown of the Beverloo scaling}

\author{Marcos A. Madrid\inst{1} \and J. R. Darias\inst{2} \and Luis. A. Pugnaloni\inst{1}}
\shortauthor{M. A. Madrid \etal}

\institute{                    
  \inst{1} Departamento de Ingenier\'ia Mec\'anica, Facultad Regional La Plata, Universidad Tecnol\'ogica Nacional, CONICET, Av. 60 Esq. 124, 1900 La Plata, Argentina.\\
  \inst{2} Laboratorio de \'Optica y Fluidos, Universidad Sim\'on Bol\'ivar, Apartado Postal 89000, Caracas 1080-A, Venezuela.
}

\abstract{
The Beverloo scaling for the gravity flow of granular materials through orifices has two distinct universal features. On the one hand, the flow rate is independent of the height of the granular column. On the other hand, less well-known yet more striking, the flow rate is fairly insensitive to the material properties of the grains (density, Young's modulus, friction coefficient, etc.). We show that both universal features are lost if work is done on the system at a high rate. In contrast to viscous fluids, the flow rate increases during discharge if a constant pressure is applied to the free surface of a granular column. Moreover, the flow rate becomes sensitive to the material properties. Nevertheless, a new universal feature emerges: the dissipated power scaled by the mean pressure and the flow rate follows a master curve for forced and unforced conditions and for all material properties studied. We show that this feature can be explained if the granular flow in the silo is assumed to be a quasistatic shear flow under the $\mu(I)$-rheology.}

\begin{document}

\maketitle



\section{Introduction}

The flow of granular matter (bulk solids such as sand, seeds, pellets, etc.) presents rather peculiar features when compared with viscous fluid flows. Since we lack a theoretical framework that can be applied to all phenomena observed in granular matter, we are still gaining knowledge from focusing on specific experimental observations and pushing to the limit our yet incomplete models. One archetypal phenomenon in this sense is the discharge of granular materials through an orifice at the bottom of a silo. The most salient feature is the fact that the flow rate does not depend on the height of the column of grains in the container, in clear contrast with the behavior of viscous fluids (see for example \cite{kadanoff1999,degennes1999,Jaeger1996,duran2000,tighe2007,beverloo,NeddermanBook} and references therein). Also striking, yet little mentioned, the particle flow rate{, $Q$,} is not affected by the material the grains are made of \cite{beverloo}. {Here, it is important to distinguish the particle flow rate ---number of particles discharged per unit time--- from the mass flow rate ---mass discharged per unit time. The latter is in fact proportional to the density of the material of the grains.}

If the discharge orifice is circular, the {particle flow rate $Q$} is described by the Beverloo rule \cite{beverloo,BrownBook} 

\begin{equation}
{Q} = C\frac{\rho_\mathrm{b}}{{m}}\sqrt{g}(D-k\,d)^{5/2}, \label{beverloo}
\end{equation}
where $D$ is the diameter of the opening, $\rho_\mathrm{b}$ the bulk density of the granular sample{, $m$ is the mass of one grain}, $g$ the acceleration of gravity and $d$ the diameter of the grains. {Note that $\rho_\mathrm{b}/m$ is the number of particles per unit volume, which is independent of the density of the material of the grains.} Here, $k$ and $C$ are two fitting dimensionless constants. Interestingly, while $k$ depends on the shape of the grains ($1.4<k<3.0$), $C$ is the same for virtually any material tested ($C\approx0.58$). Beverloo et al. \cite{beverloo}, Kondic \cite{kondic} and Mankoc et al. \cite{Mankoc2007} have stressed this ``universality'' of $C$, which is not a fundamental law of nature but a remarkable emerging feature still little investigated. 

{It is worth mentioning that the flow of grains through small orifices may clog \cite{Zuriguel2003}. However, during the short periods of flowing, the flow rate is still compatible with the Beverloo equation \cite{Mankoc2007,Thomas2013}.}
 
The standard explanation for the 5/2 power in the Beverloo rule is based on a heuristic assumption called \emph{free fall arch} \cite{NeddermanBook}. It states that most grains lose contact with the rest of the granular packing when they are about one orifice radius, $D/2$, above the outlet and that particles at this point fall freely from an initial zero vertical velocity. Then, the free fall of these grains over a distance $D/2$ leads to a flow rate proportional to $\sqrt{g} D^{5/2}$, irrespective of their material properties and column height. {Often, the argument of a constant bottom pressure in the silo is used to explain the constant flow rate (inspired by the Janssen effect observed in static silos \cite{NeddermanBook}). However, the pressure at the bottom of the silo does not remain constant throughout the discharge; rather, it falls monotonically while the flow rate remains constant \cite{aguirre2010,aguirre2011}.}

In this work, we show that adding an overweight on top of the granular column breaks down the universal features described above. {A preliminary study has shown that, when forced, the flow rate increases during the discharge (in contrast to viscous fluids) \cite{peng2009}. More interestingly, we also find here that under forcing the material properties of the grains matters.} However, for different materials and forcing conditions, the dissipated power scaled by the mean internal pressure in the silo and by the flow rate is consistent with the assumptions made for a quasistatic shear flow in the so called $\mu(I)$-rheology, based on the introduction of the inertial number $I$ \cite{dacruz}. The analysis of these extreme conditions of discharge allows us to put forward some ideas that help to understand the limits of the \emph{free fall arch} model \cite{NeddermanBook} and the universal features of the unforced discharges.

\section{Experiments}
The experimental setup is shown in fig. \ref{experiments}(a). The silo is a cylindrical glass tube (300 $\pm$ 1) mm
tall and (40.0 $\pm$ 0.5) mm in internal diameter. The bottom of the tube is bonded to an aluminum base that has a central orifice (15.0 $\pm$ 0.5) mm in diameter. The silo is filled to a height (190 $\pm$ 1) mm by pouring uniformly glass beads (glass density $2500$ kg/m$^3$) with diameter (1.00 $\pm$ 0.05) mm and bulk density {(1470 $\pm$ 20) kg/m$^3$}. We introduce a solid cylinder (piston) of Plexiglas (37 $\pm$ 1) mm tall, (39.50 $\pm$ 0.05) mm in diameter and mass (130 $\pm$ 1) g, which serves as a support for an extra overweight that can be placed on top. The downward motion of the piston was recorded using a digital camera (Pixelink PL-B741F) at 25 fps. The images were analyzed using a commercial software to obtain the height $h$ of the column of grains as a function of time. This allow us to calculate $Q(t)$ under the approximation that the packing fraction is roughly constant during the discharge. The internal friction coefficient of the glass beads is $\mu = 0.40 \pm 0.03$. 

\section{Simulations}
\begin{figure}[t]
\includegraphics[width=0.35\columnwidth, trim=0 -0.4cm 0 0, clip]{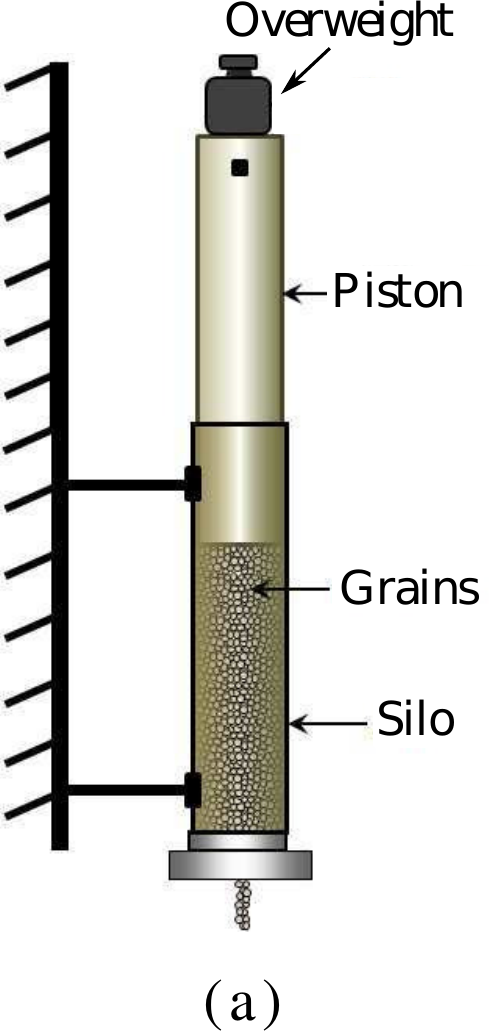}
\includegraphics[width=0.55\columnwidth, trim=0.0cm 0.0cm 0.0cm 0.0cm, clip]{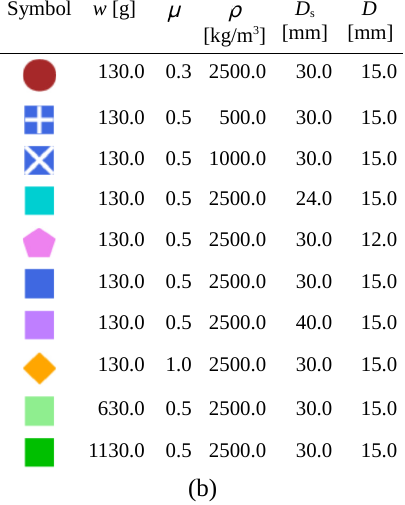}
\caption {(Color online) (a) Sketch of the experimental setup. (b) Table of parameters varied in the simulations. For each case showed there is also an unforced discharge counterpart represented in the figures as open symbols.} \label{experiments}
\end{figure}

We use Discrete Element Method (DEM) simulations via the LIGGGHTS \cite{liggghts} implementation with a particle--particle Hertz interaction and Coulomb criterion using a Young modulus $Y=70$ MPa, Poison ratio $\nu=0.25$, restitution coefficient $e=0.95$ and friction coefficient $0.3 <\mu < 1.0$ \cite{liggghts}. The same interaction applies for the particle--walls contacts. Particles are spherical with diameter $d=1$ mm. For the forced flow simulations, we introduce as an overweight a cylindrical piston. Different material densities $\rho$, silo diameters $D_\mathrm{s}=2R_\mathrm{s}$, orifice diameters $D$ and overweights $w$, are explored as summarized in fig. \ref{experiments}(b). Particles are poured to fill a height in the silo above $7 D_\mathrm{s}$ (which implies up to $2 \times 10^5$ grains, depending on the silo diameter). The orifice is initially blocked by a plug. After the grains come to rest in the silo (we wait until the kinetic energy per particle falls below $10^{-10}$ J), we remove the plug and allow its discharge. The acceleration of gravity is $g=9.81$ m/s$^{2}$. 

\begin{figure}[t]
\includegraphics[width=0.9\columnwidth, trim=.2cm 0.5cm 1.4cm 0.1cm, clip]{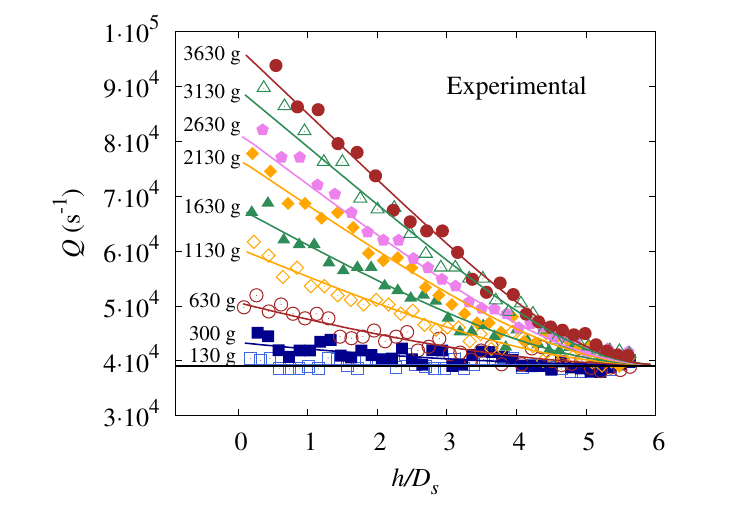}\\
\caption {(Color online) Particle flow rate $Q$ as a function of the column height in the silo measured in the experiments during the forced discharge of glass beads for different overweights (see labels in the figure). The full lines are only to guide the eye. Each curve corresponds to the average over six realizations of the experiment. {The horizontal black line corresponds to the flow rate predicted by eq. (\ref{beverloo}) with $C=0.58$ and $k=1.9$.}} \label{experiments2}
\end{figure}

\section{Results}
Figure \ref{experiments2} shows the experimental particle flow rate $Q(t)$ for forced discharges using different overweights. We have removed from the analysis the short transients at the beginning and end of the discharge. As we can see, for light overweights, $Q(t)$ is constant during the initial part and then grows in the final stages of the discharge. For heavier overweight, the acceleration of $Q(t)$ is more significant and starts earlier. This is in clear contrast to viscous fluids, which decrease in flow rate as the column empties even with an external constant forcing. The well-known rule that granular flow discharges do not depend on column height is therefore invalid for forced flows. This is a striking result. Also unforced granular discharges seem to present a very subtle increase in flow rate at the very end of the discharge \cite{geminard-private,Koivisto2016}, although no systematic study of the phenomenon has been carried out. Wilson et al. showed a similar growth in flow rate for a submerged silo (interestingly, without the need of extra overweight) \cite{wilson2014}; however, this was latter proven to be due to a hydrodynamic effect \cite{Koivisto2016}. One may speculate that the overweight induces a higher pressure at the bottom of the silo (as shown by others \cite{Ovarlez2005}) and this is responsible for the increased flow rate. We will discuss below that this cannot explain the effect. Importantly, the increase of $Q$ suggests that forced discharges cannot be explained by the \emph{free fall arch} assumption and that properties related to the specific material of the grains may become relevant. Since we find difficult varying material properties in the experiments in a controlled way, we turn in the rest of the discussion to results obtained via the DEM simulations.

Before discussing the simulation results on forced discharges, we recall here two aspects of  unforced discharges. On the one hand, as we mentioned, the particle flow rate $Q(t)$ is constant throughout the discharge and its value is mostly independent of the material properties \cite{beverloo,NeddermanBook,Mankoc2007,kondic}. However, for $\mu < 0.4$, $Q$ is reported to be slightly dependent on friction \cite{kondic}. On the other hand, the pressure at the bottom of the silo ($P_\mathrm{bottom}\equiv\sigma_{zz}(z=0)$) decays monotonically throughout the discharge \cite{aguirre2010,aguirre2011}. Results from our simulations of unforced discharges are consistent with these reported effects.

\begin{figure}[t]
\centering
  \includegraphics[width=0.8\columnwidth]{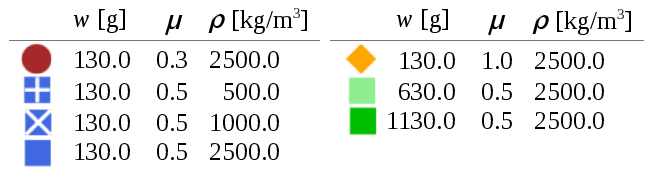}
  \includegraphics[width=0.99\columnwidth, trim=0cm 1.2cm 1.8cm 0.8cm, clip]{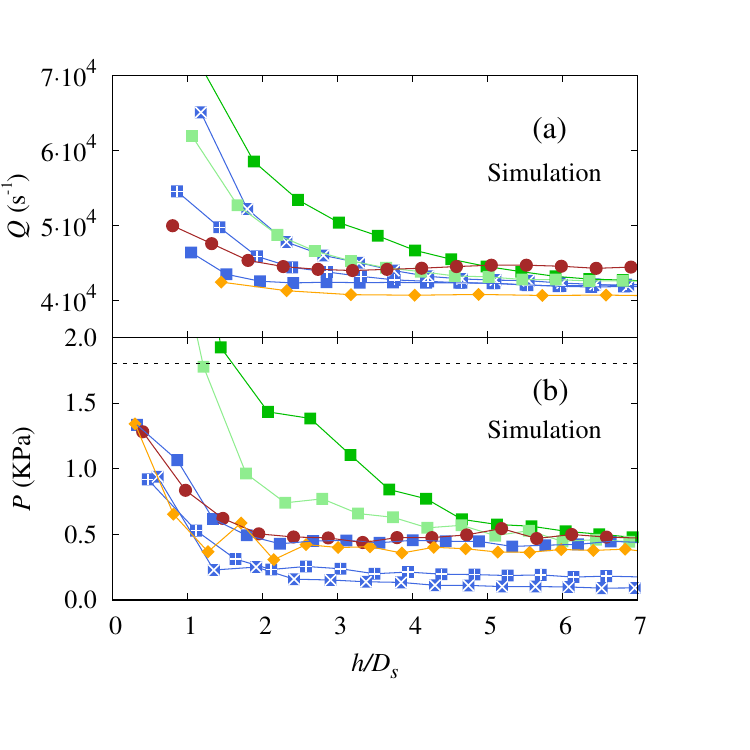}
\caption {(Color online) (a) Particle flow rate $Q$ and (b) bottom pressure $P_\mathrm{bottom}$ as a function of the height of the granular column in the simulated silo discharges {for $D_s= 30$ mm and $D=15$ mm}. Each curve corresponds to a different material and overweight. {As a reference, the horizontal dotted line in panel (b) corresponds to the pressure exerted by the $130$ g overweight.}} \label{flow-simul}
\end{figure}

In fig. \ref{flow-simul}(a), we show the simulation results for the flow rate during the discharge of spheres forced with an overweight. The results are consistent with the glass beads experiments (see fig. \ref{experiments2}), showing an increase of $Q$ towards the end of the discharge, which is more significant for heavier overweights. Figure \ref{flow-simul}(b) shows that the bottom pressure increases during the discharge since the portion of granular column screening the overweight decreases.

Interestingly, the actual increase of $Q$ in a forced discharge depends on the material properties of the grains. For example, fig. \ref{flow-simul}(a) shows that denser materials present a lower $Q$ during the acceleration phase. This seems counterintuitive since the heavier grains induce in fact a higher bottom pressure (see fig. \ref{flow-simul}(b)). We will put forward an explanation for this effect below. The friction coefficient also affects $Q$, although to a lesser extent. The larger is $\mu$, the lower is the increase of $Q$ (compare filled orange diamonds with filled blue squares in fig. \ref{flow-simul}(a)). We conclude that the universal character of the Beverloo rule, in which the particle flow rate is insensitive to the material properties, breaks down in forced flows.

As we mentioned above, one may infer that the increased pressure on the base due to the piston is the ultimate responsible for the increase in $Q$. However, bottom pressures higher than those observed using the overweight can be achieved also in an unforced discharge either by using a wider silo or by discharging materials of very high densities. All these unforced discharges yield the same low $Q$ value compared with forced discharges. Therefore, although the bottom pressure is a relevant parameter of the problem, it is clear that it does not control the flow rate on its own \cite{footnote3}.

\section{Dissipated power}

We show here that despite the different behaviors, forced and unforced silo discharges can be connected {through the rheological theory for dense granular flows, making an analogy with simpler geometries}, where valuable scalings have been found. It has been shown that for a planar shear cell of thickness $L$, {that confines a granular material between two plates with a pressure $P$, the tangential stress $\tau$, necessary to develop a shear flow at velocity $v$ can be expressed in terms of the inertial number $I$ as $\tau = \mu(I)P$ \cite{dacruz}}. {Here, $\mu(I)$ is the effective friction coefficient, that takes into account all sources of energy dissipation in the system}, $I=\frac{v}{L}\frac{d}{\sqrt{P/\rho}}$ is the inertial number that characterizes the flow (if the gains are stiff and $L\gg d$) and $\rho$ is the grains material density. {This relation between $\tau$ and $P$ through the macroscopic coefficient $\mu(I)$, summarizes the changes in the stress distribution and the exchange of linear momentum within the granular material during flow. The dissipated power $W_\mathrm{D}$ in this case is simply $W_\mathrm{D}=\tau Av$, where $A$ is the area of the  moving plate}. Jop \textit{et al}. have shown that if the grain--grain friction coefficient is above $0.4$, $\mu(I)$ is well represented by the constitutive law $\mu(I)= \mu_0 + (\mu_1 - \mu_0)/(1+I_0/I)$, with $\mu_0$, $\mu_1$ and $I_0$ three constants which are independent of the material properties of the grains \cite{Jop2006}.

{In the case of the silo, we can compare the flow during discharge with the flow in a planar shear cell by considering the shear stress $\tau$ exerted on the granular material. In this case, the grains flow exerting pressure against the walls of the silo thanks to Reynolds dilatancy, giving rise to an increase of the confining mean stress $P$ in the column. The stationary flow pattern developed can be characterized by the shear rate $v/L$, where $v$, is the velocity of the free surface and $L$ is the silo radius \cite{footnote1}. Then, we can write the power, $W_\mathrm{D}$, dissipated by the granular material during flow at any given time as}

\begin{equation}
 W_\mathrm{D}(t)= \tau(t) A(t) v(t) = \mu(I) P(t) A(t) \frac{Q(t)m}{\rho_\mathrm{b} A_\mathrm{s}}, \label{dissipation}
\end{equation}
where $A(t)$ is the area of frictional contact between the grains and the silo, $A_\mathrm{s}$ is the cross section of the silo, and we have used the continuity relation $m Q(t)= \rho_\mathrm{b} A_\mathrm{s} v(t)$. In this equation, $A(t)$ includes the contact with the lateral walls, which decreases during the discharge, plus a fixed area accounting for the effect of the converging flow at the lower portion of the silo, which is expected to be proportional to the base of the silo. Hence, we can express $A(t)$ as

\begin{equation}
A(t)=2\pi R_\mathrm{s} h(t)+\alpha\pi R_\mathrm{s}^2, \label{a-t}
\end{equation}
where $h(t)$ is the height of the granular column at time $t$, and $\alpha$ is a constant. 

For wide silos, $L\rightarrow \infty$ and hence $I\rightarrow 0$. In our simulations with a finite diameter silo the inertial number varies during the discharge in the range $9.8\times 10^{-3} <I< 1.6 \times 10^{-2}$, which is mostly in the quasistatic limit ($I\lesssim10^{-2}$) \cite{dacruz}. {In the quasistatic limit, $\mu(I)\rightarrow \mu_0$, and does not depend on the material \cite{Jop2006}. For simple shear simulations in 2D $\mu_0 \approx 0.26$ \cite{dacruz} and in 3D $\mu_0\approx 0.35$ \cite{azema2014}.} Therefore, from eqs. (\ref{dissipation}) and (\ref{a-t}), 
 
\begin{equation}
W_\mathrm{{D}}^*(t) = \frac{\rho_\mathrm{b}}{m}\frac{W_\mathrm{D}(t)}{P(t)Q(t)}=\mu_0\frac{A(t)}{A_\mathrm{s}}=\mu_0  \left[ \frac{4 h(t)}{D_\mathrm{s}} + \alpha \right], \label{dissipation2}
\end{equation}
which is a quantity that should not depend on the material properties nor orifice size $D$. Note that $\rho_\mathrm{b}/m$ is the number density, which does not depend on the grains material density. 

\begin{figure}[t]
\includegraphics[width=\columnwidth]{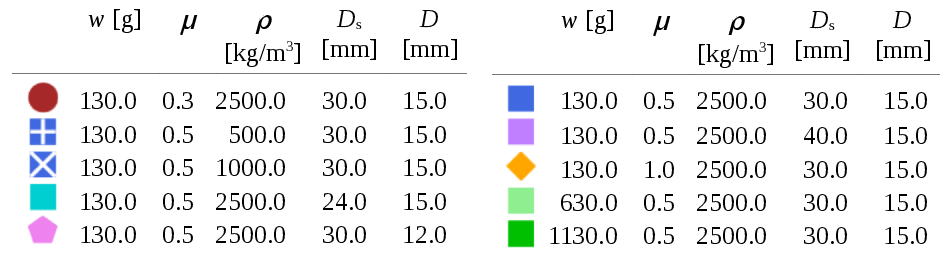}
\includegraphics[width=\columnwidth, trim=0.0cm 0.1cm 0.5cm 0.2cm, clip]{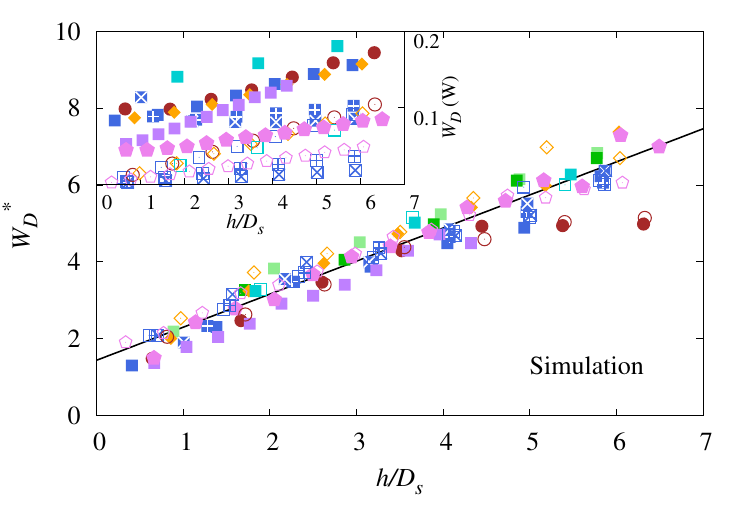}
\caption {(Color online) $W_\mathrm{{D}}^*(t) = \rho_\mathrm{b}W_\mathrm{D} / (m P Q)$ during a discharge as function of the column height $h$ scaled by the silo diameter for different $\rho$, $\mu$, $D$, $D_{\rm s}$ and overweights [see fig. \ref{experiments}(b) for reference]. Filled symbols correspond to forced discharges and open symbols to corresponding unforced discharges. The solid line corresponds to the fit using eq. (\ref{dissipation2}). The inset shows $W_\mathrm{D}$ before scaling by $PQ$.} \label{mastercurve}
\end{figure}

Figure \ref{mastercurve} shows $W_\mathrm{{D}}^*$ as a function of the column height scaled by $D_s$ for a series of simulations: forced and unforced, and for various values of $\rho$, $\mu$ and $D_{\rm s}$. The data is consistent with eq. (\ref{dissipation2}) by fitting $\mu_0=0.21 \pm 0.01$ and $\alpha = 6.7 \pm 0.6$. We have also included results for a discharge with an orifice of $D=12$ mm (see pink pentagons in fig. \ref{mastercurve}). We notice that the low friction sample ($\mu=0.3$, red circles) deviates from the set of data. As we mentioned, this is expected since $\mu_0$ was shown to depend on friction if $\mu < 0.4$ \cite{dacruz}. The value of $\mu_0$ found is different from the values obtained in planar shear flows in 2D and 3D. This indicates that geometry is a factor that affects the effective friction of the material. One should expect then that $\mu_0$ will vary with silo geometry. The relatively high value of $\alpha$ obtained indicates that a significant part of the energy is in fact dissipated in the converging region of the flow, where shear rate is more important.

{It is important to notice that the dissipated power is not proportional to the grain--wall friction. The inset to fig. \ref{mastercurve} shows that the dissipated power is the same for systems where only $\mu$ has been changed (solid blue squares, solid orange diamonds, solid red circles). This is so because the internal pressure [fig. \ref{flow-simul}(b)] is the same for different $\mu$ and the ``effective friction'' [i.e., $\mu(I)$] is also the same. Notice however that the case for $\mu=0.3$ deviates a slightly from since below $\mu=0.4$ one expect a dependency of $\mu(I)$ on $\mu$.}

The scaling shown by fig. \ref{mastercurve} demonstrates that the flow in a silo discharge could be modeled as a quasistatic shear flow, for unforced and forced conditions, without the need of heuristic postulates such as the \emph{free fall arch}. A first modeling attempt in this direction can be found in ref. \cite{madrid2016}.

\section{Discussion}
From a ``microscopic'' perspective, the negligible effect of the material properties of the grains on the flow rate for unforced discharges can be understood as follows. Since the collision rate observed in a dense granular media
is very high, all energy input (due to gravity) is dissipated in a very short time. Then, the discharging granular column cannot accelerate because the energy input cannot be converted into kinetic energy. Increasing the material density will increase the power injected by gravity but also the friction force at each contact will increase proportionally (heavier grains induce larger normal contact forces). As a consequence, the system will still be able to dissipate all injected power and avoid acceleration. Reducing the friction coefficient may change the dissipative capacity of each collision; however, the number of collisions is so large that all energy input will still be dissipated. An exception to this can be found only if the dissipative properties of the material are very poor (e.g., $\mu<0.4$) \cite{kondic} or if the number of grains in the silo is very small (at the end of the discharge) \cite{Koivisto2016}.  

When a piston does work on the system at a high rate, the collisions are unable to fully dissipate the energy input, particularly when the number of particles in the column decreases. As a result, the flow rate increases due to the non-dissipated kinetic energy.

In forced flows, the increase of flow rate does depend on the dissipative properties of the grains. We showed that a higher $\mu$ leads to a lower flow rate increase [see fig. \ref{flow-simul}(a)]. More interestingly, heavy grains display an intriguing low increase in $Q$. To explain this, let us consider the simulations with densities $1000$ kg/m$^3$ and $2500$ kg/m$^3$ forced with a $130.0$ g overweight [filled blue squares with and without pluses, respectively, in fig. \ref{flow-simul}]. On the one hand, the power injected initially into the system is proportional to the total weight (grains plus piston) which is $\approx 200$ g for the light grains and $\approx 320$ g for the heavy grains. On the other hand, the dissipation at each frictional contact is proportional to the internal pressure, which is initially $130$\% higher for the heavier grains [see fig. \ref{flow-simul}(b)]. Hence, with this overweight, doubling the material density leads to a $60$\% increase in injected power, but to a $130$\% increase in dissipative capacity. As a consequence, the heavier grains are able to dissipate a larger proportion of the injected power and so induce a smaller increase in $Q$.

\section{Conclusions}
By injecting energy into a discharging silo at a high rate, we have shown that not only the independence of granular flow rate with the system height is broken, but also that the universality with respect to material properties is lost. In contrast to viscous fluids, the flow rate is accelerated during the discharge.

The flow can be described as a shear flow consistent with the $\mu(I)$-rheology in the quasistatic limit. As a consequence, the dissipated power scaled by the mean pressure and the flow rate becomes rather independent of the material properties of the grains if the grain-grain friction coefficient is above 0.4. This opens the opportunity for modeling silo discharges in a wide range of conditions without the need of heuristic approximations such as the \emph{free fall arch} and the \emph{empty annulus}, which have been recently challenged \cite{janda2012,rubio2015}.

\acknowledgments
We acknowledge valuable discussions with D. Maza, L. Kondic, I. Zuriguel, J-C. G\'eminard, M. A. Aguirre, D. Durian and E. Cl\'ement. This work has been supported by ANPCyT (Argentina) through grant PICT-2012-2155, Universidad Tecnol\'ogica Nacional (Argentina) through grant  PID-MA0FALP0002184, Centro Argentino Franc\'es de Ciencias de la Ingenier\'ia (CAFCI, Argentina-Francia) and FONACIT through grant 2015000072 (INVUNI2013-1563) Universidad Sim\'on Bol\'ivar (Venezuela).


\begin{thebibliography}{0}
\bibitem{footnote3} This is in line with a remark by Staron et al. \cite{Staron} that show that the pressure at the silo base can be changed by using silos of different diameters, however the flow rate remains the same.
\bibitem{footnote1} The pressure in a granular column varies significantly as a function of the vertical coordinate. Also the shear rate (estimated by measuring the difference in mean velocity of the grains at the central axis and the grains close to the walls and dividing by the silo radius) is different at different heights during a discharge.  For a flat bottomed silo, discharging within a mass flow regime, close to the free surface the shear rate is negligible. However, close to the base, the shear rate is maximum since the velocity at the walls is zero (due to the stagnant zone) and the velocity at the outlet is the highest. Taking the velocity at the free surface and averaging the pressure over the entire column is a rather crude approximation. However, the scaling found in this work is already good. Improvements can be made by taking a layer-wise approach.
\end{thebibliography}
\end{document}